\documentclass{iaus}
\usepackage{graphicx}

  \checkfont{eurm10}
  \iffontfound
    \IfFileExists{upmath.sty}
      {\typeout{^^JFound AMS Euler Roman fonts on the system,
                   using the 'upmath' package.^^J}%
       \usepackage{upmath}}
      {\typeout{^^JFound AMS Euler Roman fonts on the system, but you
                   dont seem to have the}%
       \typeout{'upmath' package installed. iaus.cls can take advantage
                 of these fonts,^^Jif you use 'upmath' package.^^J}%
      }
  \else
  \fi

  \checkfont{msam10}
  \iffontfound
    \IfFileExists{amssymb.sty}
      {\typeout{^^JFound AMS Symbol fonts on the system, using the
                'amssymb' package.^^J}%
       \usepackage{amssymb}%

      }{}
  \fi

  \IfFileExists{amsbsy.sty}
    {\typeout{^^JFound the 'amsbsy' package on the system, using it.^^J}%
     \usepackage{amsbsy}}
    {}

\def\aj{AJ}%
\def\apj{ApJ}%
\def\apjs{ApJS}%
\def\mnras{MNRAS}%
\def\pasp{PASP}%
\def\pasj{PASJ}%
\newsavebox{\astrutbox}
\sbox{\astrutbox}{\rule[-5pt]{0pt}{20pt}}

\newcommand\etal{\mbox{\textit{et al.}}}

\title[Outskirts of Galaxy Clusters: intense life in the suburbs]
      {Intracluster light at $z\sim0.25$ from SDSS imaging data}

\author[S. Zibetti, and S. D. M. White]
{Stefano Zibetti$^1$
\and
Simon D. M. White$^1$}

\affiliation{$^1$Max-Planck-Institut f\"ur Astrophysik, Garching bei M\"unchen,
Germany \\ email: zibetti@mpa-garching.mpg.de}

\pubyear{2004}
\volume{195}
\pagerange{1--5}
\date{March 12th, 2004 and in revised form ??}
\jname{Outskirts of Galaxy Clusters: intense life in the suburbs}
\editors{A. Diaferio, ed.}
\begin{document}

\maketitle

\begin{abstract}
We investigate the broadband optical emission of diffuse intergalactic stars
in galaxy clusters at $z\sim0.25$
by means of an image stacking technique. The images of 654 clusters,
selected with the max-BCG algorithm from a subsample of the SDSS-Data Release 1 (DR1),
have been stacked after masking all the sources detected down to very low S/N.
The resulting images in the g, r and i bands provide reliable photometric data at the 
level of $\gtrsim30$~mag$/\square$" (r band), out to $\gtrsim 600$~kpc from the 
Brightest Cluster Galaxy (BCG).
\\Our analysis shows that:\\
i) the IntraCluster Light (ICL) is much more concentrated than the galaxy light,
contributing $\sim$30\% of the total cluster optical emission at $R=100$~kpc, but
less than 10\% at $R>500$~kpc;\\
ii) the ICL contributes between 15 and 20\% of the total cluster optical luminosity
between the optical radius ($\mu_r=25$~mag$/\square$") of the BCG and 500 kpc;\\
iii) the colours of the ICL are consistent with the global colours of the cluster galaxies,
with little evidence for redder g-r.
\end{abstract}

\firstsection 
\section{Introduction}

Firstly proposed by \cite{1951PASP...63...61Z}, the presence of a diffuse population of intergalactic 
stars in galaxy clusters is now a well established observational fact 
(see e.g. \cite[Arnaboldi 2003]{2003IAUS..217E..20A} and references therein, Krick \& Bernstein this conference).
Such a population is commonly believed to 
originate from the disruption of small galaxies and from the tidal stripping of the
outskirts of large galaxies in high-density environments. Studying the properties
of the ICL as a function of redshift and of the richness and morphology of the parent cluster
can give a significant insight in the interaction processes that drive galaxy
evolution in dense environments.\\
Observing the intracluster light (ICL) is extremely challenging, due to the
very low surface brightness involved ($>26$~mag$/\square$"), which is less
than 1\% of the typical surface brightness of the sky. 
During the last years the increasing sensitivity of CCD detectors and the
development of new observational techniques has made it possible to start a number of
optical surveys aiming at the study of the ICL,
by means of both broadband imaging of unresolved stellar 
populations (e.g. \cite[Feldmeier \etal~2002]{2002ApJ...575..779F}) and 
of narrow band detection of intracluster planetary nebulae (e.g.
\cite[Arnaboldi \etal~2003]{2003AJ....125..514A}). 
Nevertheless, for a statistically representative number of clusters of different 
morphology and richness and located at different redshifts 
we still lack reliable determinations
of the basic properties of the ICL, such as the fractional amount with respect
to the total optical emission of the cluster, the radial distribution,
the link with (possible) cD envelopes, or the colours.
In fact, observations of the ICL in individual clusters are extremely demanding in
telescope time and require an incredibly high control of the flat fielding and of the
sources of light pollution, such as fore- and background sources and the internal reflections
of the camera.\\
We have chosen to tackle the problem with a purely statistical 
approach,
adopting an image stacking technique, similar to that used by \cite{2004MNRAS.347..556Z}.
The Sloan Digital Sky Survey (SDSS, \cite[York \etal~2000]{2000AJ....120.1579Y}) is providing 
a one-quarter-of-sky coverage with 5-band imaging 
(\cite[Fukugita \etal~1996, Gunn \etal~1998]{1996AJ....111.1748F,1998AJ....116.3040G})
and spectroscopy. From a subsample of its first data release 
(DR1, \cite[Abazajian \etal~2003]{2003AJ....126.2081A}),
we have extracted a catalogue of 654 clusters and rich groups in the redshift range 0.2-0.3, 
whose imaging in the g, r, and i bands was processed and stacked, as 
described in Sec.\ref{stacking}.
The results are presented and discussed in Sec.\ref{results}.

\section{Method: the stacking technique}\label{stacking}
The very essence of the stacking technique consists in averaging a large number of images of
clusters, in order i) to increase the signal-to-noise ratio and measure surface brightnesses 
as faint as $\mu_r\sim30$~mag$/\square$", and ii) to statistically remove the 
effects of flat fielding inhomogeneities and the contamination from scattered light, internal
reflections, and fore- and background sources. Given the typical S/N in the SDSS imaging and
the limitation imposed by the present day coverage of the survey, a suitable number of
images ranges from several hundreds to roughly 1,000.\\

\subsection{The SDSS sample}\label{sample}
The sample selection focused on clusters in the redshift range 0.2-0.3 for two main 
reasons. First, we want the spatial extension of the individual clusters not to exceed the
dimensions of the frame, since an annulus of at least 1 Mpc around the cluster must be 
included on the same frame for a reliable background subtraction. 
Given the size of the SDSS frames,
the chosen
redshift range represents the best trade-off between this spatial constraint and the desire
for minimum cosmological dimming of the apparent surface brightness and for maximum 
efficiency in detecting galaxies. On the other hand,
at $z=0.25$ the g-r colour conveniently maps the 4000\AA-break, thus providing
important information about the intracluster stellar population.\\
The cluster sample is drawn from $\sim1500$~deg$^2$ of photometric data in the SDSS DR1,
using the maxBCG method by \cite{annis04} 
(see also \cite[Bahcall \etal~2003]{2003ApJS..148..243B} for a description of this technique).
We require estimated photometric redshift $0.2<z<0.3$, 17 or more galaxies identified as
red-sequence members within 1~Mpc projected distance\footnote{$H_0$=70km sec$^{-1}$ Mpc$^{-1}$,
$\Omega_0$=1, $\Omega_\Lambda$=0.7 throughout this paper.} from the BCG, and at least 13 
of them within 330 kpc. The clusters selected in this way have richness roughly
in the same range as the Abell clusters. Clusters in vicinity of bright foreground
stars or in frames
contaminated by scattered light have been pruned from the sample.

\subsection{Image processing}
The SDSS imaging data are available as bias subtracted, flat-field corrected frames.
For each cluster we estimate the sky background in an annulus of 1~Mpc,
100~kpc thick, centered on the BCG, after masking all the sources detected at 
very low S/N with SExtractor, in at least one of the three passbands.
After subtracting the background, for each image two kinds of masks are built, ``ICL'' and
``galaxy''. The former are obtained
from SExtractor segmentation images, adopting fixed surface brightness detection thresholds 
($\mu_r$=24.5, $\mu_g$=25.0, $\mu_i$=24.0) and minimum detection area of 10 contiguous pixels.
Corrections for galactic extinction and cosmological dimming
($(1+z)^4$) with respect to the median redshift of the sample are applied. 
The masks of bright sources are
grown by 30 pixels in order to avoid scattered light, while the BCG is left unmasked.
The final ``ICL'' mask results from the OR combination of the masks obtained for the 
three passbands. In the ``galaxy'' masks only the bright sources are masked.\\
The images and the corresponding masks are geometrically transformed, centered on the BCG,
rescaled to the same physical units and randomly rotated. Finally, the images are averaged
excluding the pixels masked according either to the ``ICL'' or to the ``galaxy'' masks, 
so that it is possible to estimate the average contribution
from the diffuse light (plus the BCG) and the average total light, respectively.

\section{Results}\label{results}
\begin{figure}
\centerline{
\includegraphics[height=5.5truecm,width=4.7truecm]{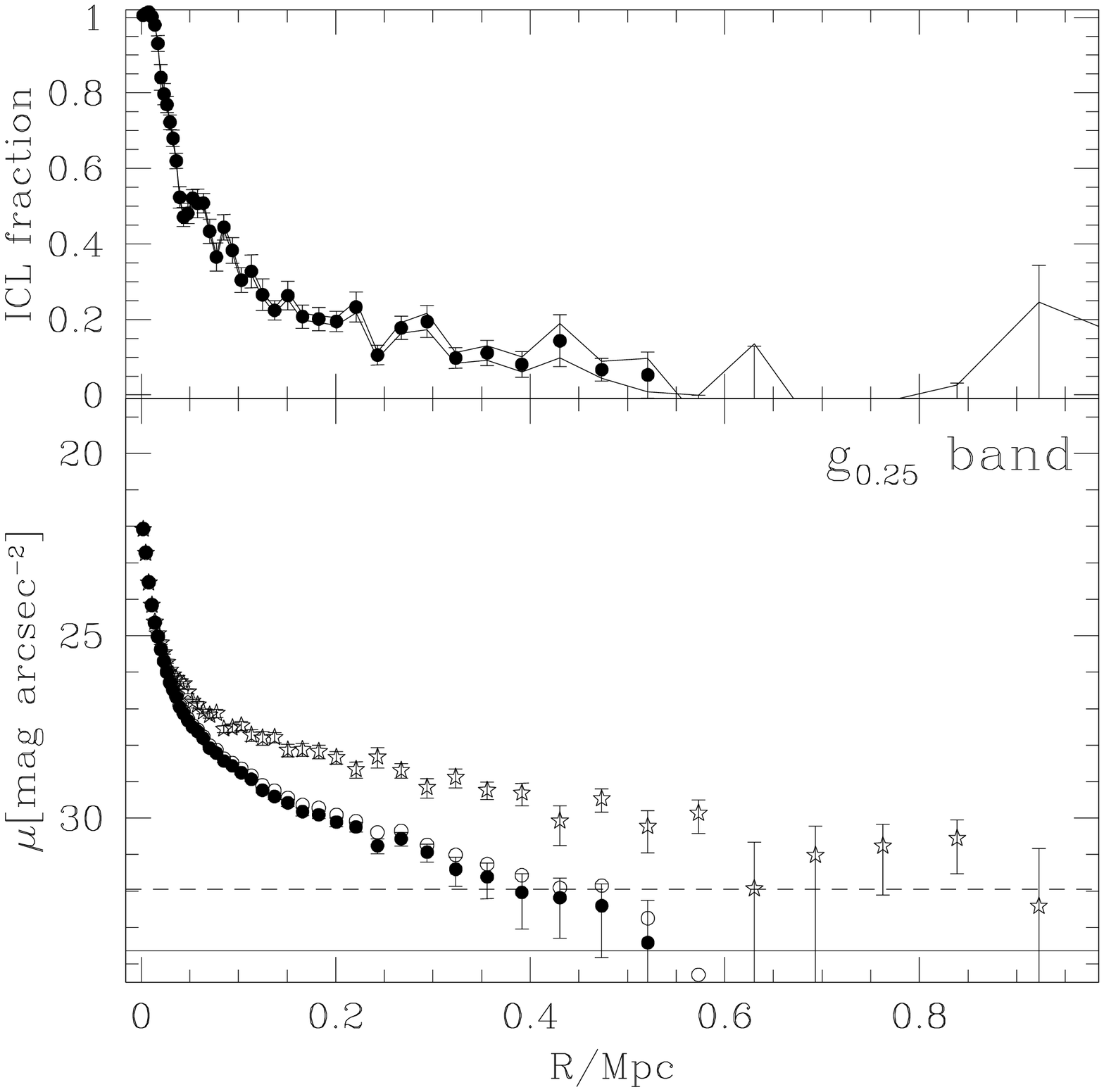}
\includegraphics[height=5.5truecm,width=4.7truecm]{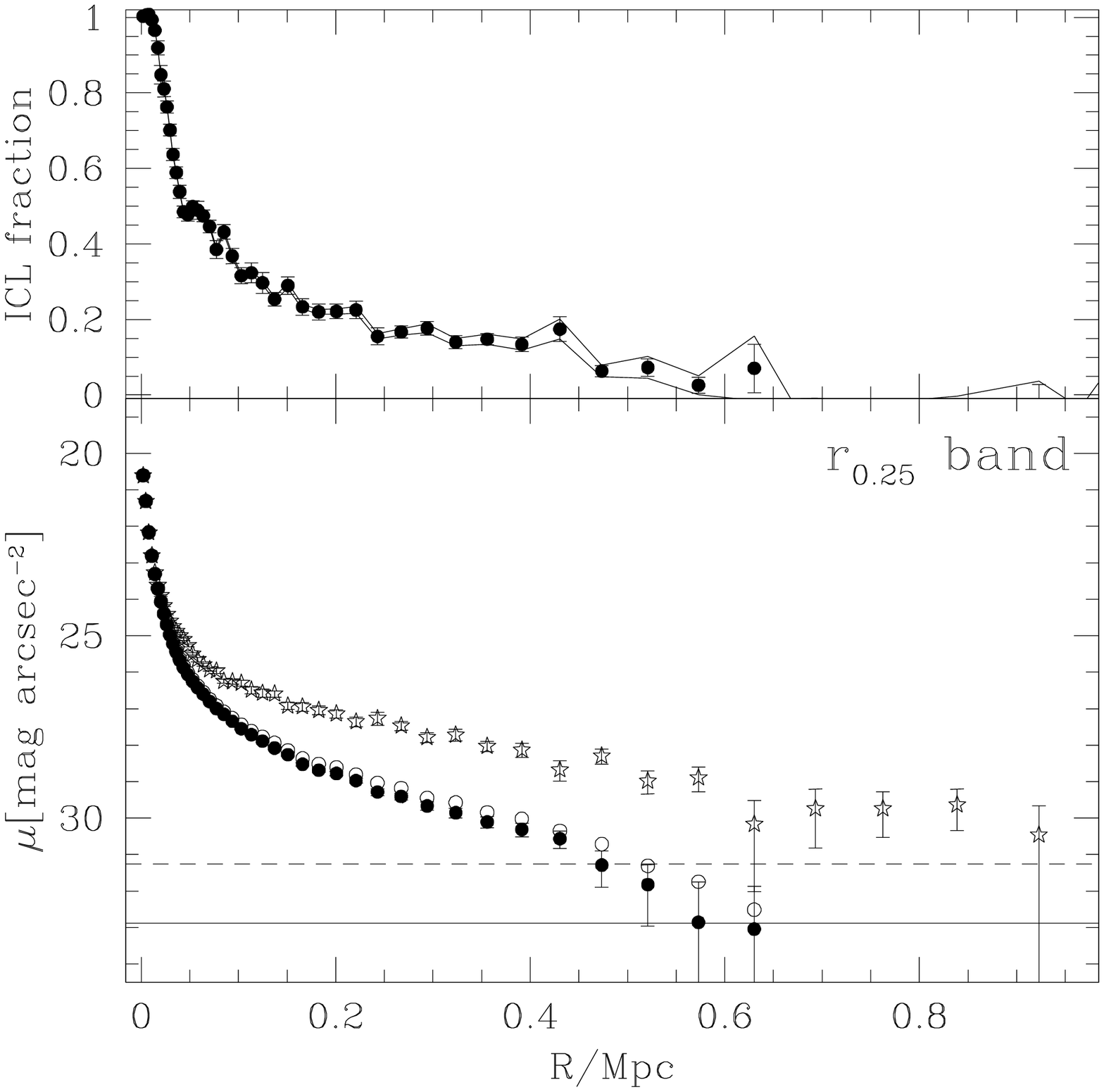}
\includegraphics[height=5.5truecm,width=4.7truecm]{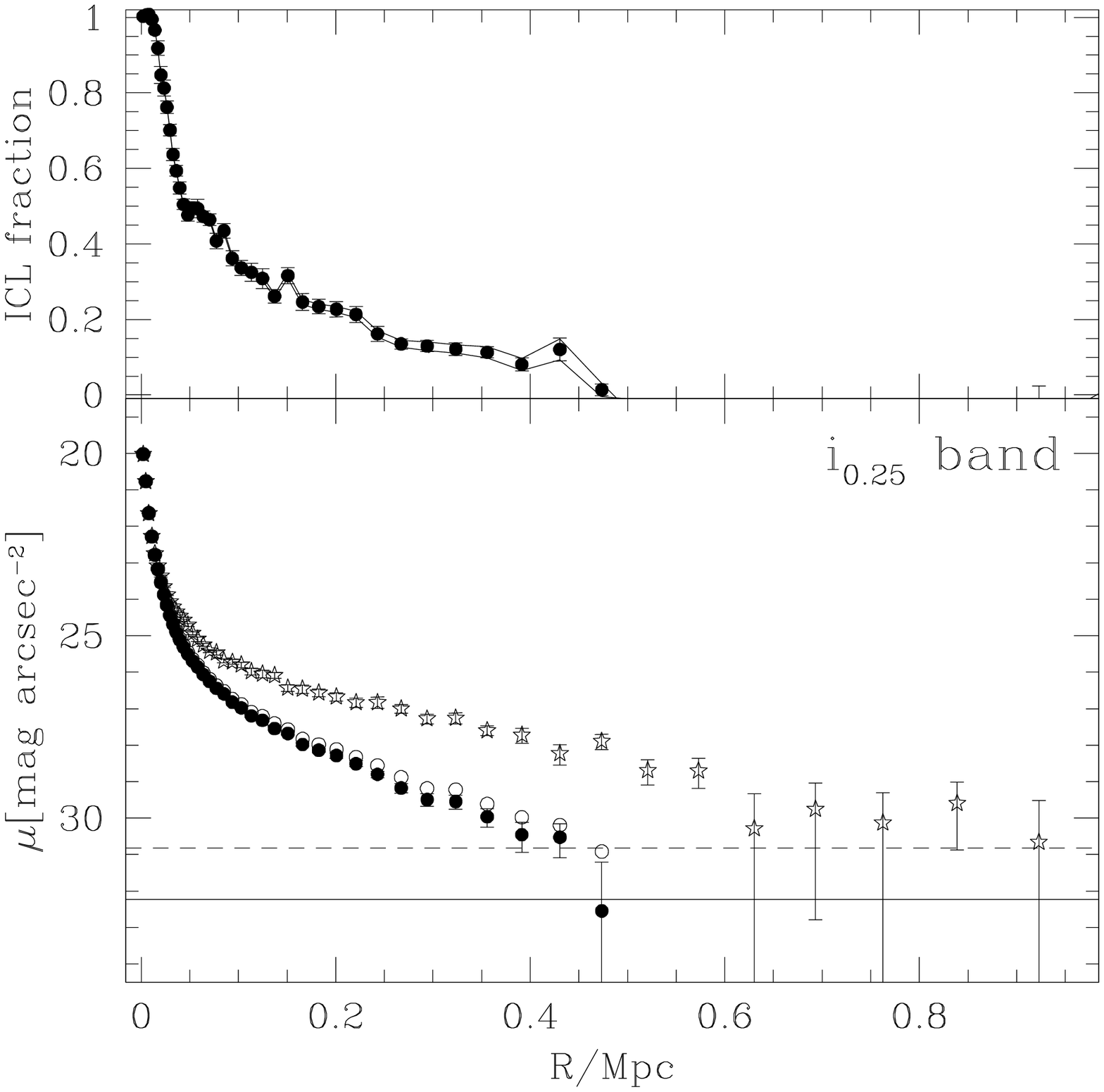}
}
\caption{Lower panels: SB profiles for the total light (stars), the measured diffuse light
(open circles) and the corrected ICL (filled circles). Lines display the background
uncertainty for the total light (dashed line) and for the diffuse component (solid line).
Errorbars are 1-$\sigma$ errors, including random noise and systematic background uncertainty.
Upper panels: the local ICL/total ratio. Errorbars account for random errors only, while the
solid lines represent the 1-$\sigma$ intervals when accounting for systematic errors in the
background estimation.}\label{profiles}
\end{figure}

\begin{figure}
\centerline{
\includegraphics[height=8.5truecm]{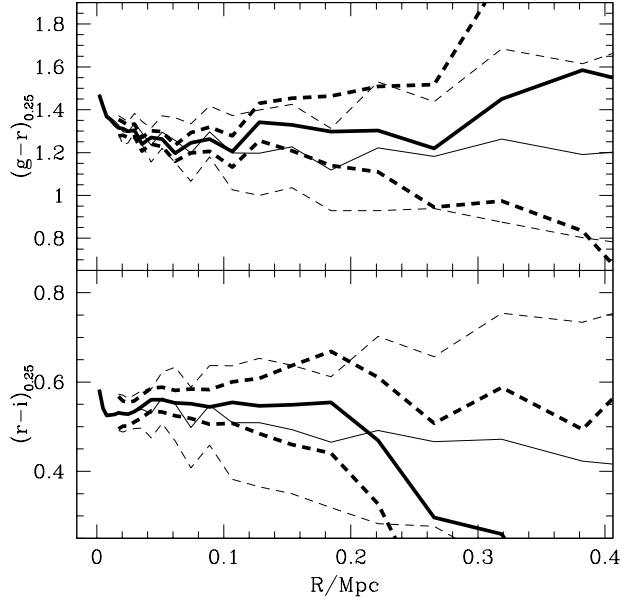}
}\caption{g-r and r-i colour profiles. Dashed lines represent the 1-$\sigma$
confidence intervals, thick lines are for the ICL, thin for the total cluster light.}\label{colors}
\end{figure}

In Fig.\ref{profiles} (lower panels) we show the azimuthally averaged SB profiles 
extracted from the stacked 
images, after subtracting the background evaluated at 1~Mpc, in g, r, and i band 
respectively. 
The star points represent the total cluster light (``galaxy'' masks),
while the open circles are the diffuse light (``ICL'' masks). The solid
and the dashed lines display the level of the background uncertainty for the diffuse component and
for the total light, respectively. As apparent from these plots, we are able to measure
SB\footnote{All surface brightnesses are expressed in the observer's frame filter system for
clusters at $z$=0.25.
} as faint as 
$\mu_r\simeq32$ for the diffuse component, and $\mu_r\simeq30$ for the total light,
the latter being limited by the shot noise of the background galaxy distribution.\\
Whether the diffuse light we measure is ``true'' ICL is questionable, depending on
the actual definition of ICL. In principle one should consider as ICL only the light coming from 
stars which are unbound from individual galaxies and freely fluctuating in the cluster potential.
Since this is clearly not a viable operational definition, we consider as ICL 
all the light coming from outside of the optical radius of galaxies (i.e. the isophotal 
radius at $\mu_r=25$). This is a well defined quantity, that can be also easily tested 
against theoretical predictions. However, due to the magnitude limits of the SDSS (m$_r$=22.2),
we expect significant contribution to the diffuse light from undetected cluster 
galaxies. We quantify this contamination by simulating a set of mock galaxy clusters, in the
same observational conditions (redshift, galactic extinction, seeing) as the observed ones.
Each simulated cluster includes 1,000 galaxies, whose apparent magnitude is assigned
according to a montecarlo realization of a Schechter luminosity function (LF). 
We use the distribution of Sersic profile parameters as a 
function of luminosity as observed 
by \cite{blanton03} to create a realistic 2D image of each galaxy, truncated at the 
$\mu_{r}=25$ isophote. The images are then convolved with the observed PSF, background and 
poissonian noise are added. After processing and stacking the mock images in the same way as the
observed ones, we measure the fraction of galaxy light that is missed by the masking
algorithm and contributes to the diffuse
component. This fraction depends on the assumed LF, and ranges from a few percent for
a flat LF, up to 15\% in the case of a steep faint end. 
Assuming the LF
given by \cite{gotoLF} ($M_r^*$=-22.21, $\alpha$=-0.85), this fraction is 5\% (r band).\\
The corrected SB of the ICL
is represented by filled circles in the lower panels of Fig.\ref{profiles}.
We find that the ICL is much more concentrated
than the total light in the cluster, irrespective of the passband. The local
fractional contribution of the ICL is shown in the upper
panels of Fig.\ref{profiles}. Note that, since we do not mask the BCG, the fraction is 1 in the
center. At 100~kpc the fraction of light contributed by the ICL
is $\sim30$\%, and drops to $\sim5$\% at 500~kpc. The overall fraction of ICL between
the average optical radius of the BCG (30~kpc) and 500~kpc
is 20\% in the r band, and 19\% in the g and i bands, with typical
uncertainties of $\pm$2\%. Assuming a steeper LF can lower these figures by up to $\sim5$\%.\\
Using the three observed bands, we study the g-r and r-i colour profiles of the ICL out to
400~kpc from the BCG. They are shown in Fig. \ref{colors} as the thick solid lines.
The total-light colours are shown as thin solid lines. Dashed lines display the $1-\sigma$
confidence intervals.
We note a strong negative colour gradient in the central 20 kpc, that is inside the BCG.
The outer profiles do not display any significant gradient in the galaxy component, although
marginally bluer r-i is measured at larger radii.
Neglecting the r-i at $R>0.2$ Mpc, whose value is largely undetermined due to the very low signal,
we find that the ICL colours are very similar to those of the galaxy component, 
with perhaps weak evidence for marginally redder g-r.\\
This result and the strong, non-linear dependence of the ICL surface brightness
on the galaxy density, are consistent with the idea of ICL being
emitted by passively evolving stellar populations, stripped from the cluster 
galaxies in the highest density regions.\\
The forthcoming analysis of a larger sample of clusters (\cite[Zibetti \etal~2004~in preparation]{ziby_prep}), differentiated
according to their richness
and morphology (e.g. Bautz-Morgan type), will contribute new clues about the origins of the ICL.

\begin{acknowledgements}
We thank Jim Annis for kindly providing us with the maxBCG catalog of clusters used in this
work.\\

Funding for the creation and distribution of the SDSS Archive has been provided by 
the Alfred P. Sloan Foundation, the Participating Institutions, the National Aeronautics 
and Space Administration, the National Science Foundation, the U.S. Department of Energy, 
the Japanese Monbukagakusho, and the Max Planck Society. The SDSS Web site is 
http://www.sdss.org/.
The SDSS is managed by the Astrophysical Research Consortium (ARC) for the Participating 
Institutions. 

\end{acknowledgements}


\begin{thebibliography}{}
\bibitem[Abazajian \etal (2003)]{2003AJ....126.2081A} {Abazajian, K.~\etal} 
2003, \textit{\aj}, \textbf{126}, 2081--2086

\bibitem[Annis \etal~(2004)]{annis04} {Annis, J.~\etal} 
2004, in preparation

\bibitem[Arnaboldi~(2003)]{2003IAUS..217E..20A} {Arnaboldi, M.} 2003, 
In \textit{Recycling intergalactic and interstellar matter} 
(Eds. P. A. Duc, J. Braine, \& E. Brinks). IAU Symposium Series, vol.\ 217 
(\textit{astro-ph/0310143})


\bibitem[Arnaboldi \etal~(2003)]{2003AJ....125..514A} {Arnaboldi, M.~\etal} 
2003, \textit{\aj}, \textbf{125}, 514--524

\bibitem[Bahcall \etal~(2003)]{2003ApJS..148..243B} {Bahcall, N.~A.~\etal} 
2003, \textit{\apjs}, \textbf{148}, 243--274

\bibitem[Blanton \etal~(2003)]{blanton03} {Blanton, M.~R.~\etal} 
2003, \textit{\apj}, \textbf{594}, 186--207

\bibitem[Goto \etal~(2002)]{gotoLF} {Goto, T.~\etal} 2002, 
\textit{\pasj}, \textbf{54}, 515--525

\bibitem[Feldmeier \etal~(2002)]{2002ApJ...575..779F} {Feldmeier, J.~J., 
Mihos, J.~C., Morrison, H.~L., Rodney, S.~A., \& Harding, P.} 2002, 
\textit{\apj}, \textbf{575}, 779--800

\bibitem[Fukugita \etal~(1996)]{1996AJ....111.1748F} {Fukugita, M., 
Ichikawa, T., Gunn, J.~E., Doi, M., Shimasaku, K., \& Schneider, D.~P.} 
1996, \textit{\aj}, \textbf{111}, 1748-+

\bibitem[Gunn \etal~(1998)]{1998AJ....116.3040G} {Gunn, J.~E., \etal} 1998, 
\textit{\aj}, \textbf{116}, 3040--3081

\bibitem[York \etal~(2000)]{2000AJ....120.1579Y} {York, D.~G.~\etal} 2000, 
\textit{\aj}, \textbf{120}, 1579--1587

\bibitem[Zibetti, White, \& Brinkmann (2004)]{2004MNRAS.347..556Z} {Zibetti, 
S., White, S.~D.~M., \& Brinkmann, J.} 2004, \textit{\mnras},
\textbf{347}, 556--568

\bibitem[Zibetti \etal~(2004 in preparation)]{ziby_prep} {Zibetti, 
S.~\etal} 2004, in preparation

\bibitem[Zwicky(1951)]{1951PASP...63...61Z} {Zwicky, F.} 1951, 
\textit{\pasp}, \textbf{63}, 61-+
\end{thebibliography}
\end{document}